\documentclass[preprint]{revtex4-1}

\usepackage{latexsym}
\usepackage{amsmath}
\usepackage{amsfonts}
\usepackage{amssymb}
\usepackage{amsbsy}
\usepackage{bm}
\usepackage{color}
\usepackage{graphicx}
\usepackage{longtable}
\usepackage{setspace}
\usepackage{color}
\usepackage{hyperref}
\usepackage{soul}

\newcommand{\on}{\omega^{o}_{\mathrm{on}}}
\newcommand{\off}{\omega_{\mathrm{off}}}

\begin{document}

\begin{titlepage}
\title{Filament extensibility and shear stiffening control persistence of strain and loss of coherence in cross-linked motor-filament assemblies}
\author{Arvind Gopinath$^{1,*,\dag}$, Raghunath Chelakkot$^{2,*}$ and  L. Mahadevan$^{3,\ddag}$}
\affiliation{$^{1}$Department of Bioengineering, University of California Merced,  Merced, CA, USA.\\
$^{2}$ Department of Physics, Indian Institute of Technology Bombay, Mumbai, India.\\
$^{3}$Department of Physics, and School of Engineering and Applied Sciences,  Harvard University, Cambridge, MA, USA. 
\\$^{*}$Equal contribution, 
$^{\dag}$agopinath@ucmerced.edu, $^{\ddag}$lmahadev@g.harvard.edu}

\begin{abstract}
Cross-linked flexible filaments deformed by active molecular motors occur in many natural and synthetic settings including eukaryotic flagella, the cytoskeleton and in vitro motor assays. In these systems, an important quantity that controls spatial coordination and emergent collective behavior is the length scale over which elastic strains persist.  We estimate this quantity in the context of ordered composites comprised of cross-linked elastic filaments sheared by active motors. Combining a mean-field theory valid for negligibly noisy systems with  discrete simulations for noisy systems, we show that the effect of localized strains -- be they steady or oscillatory --
persist over distances determined by motor kinetics, motor elasticity and filament extensibility. The cut-off length that emerges from these effects controls the transmission of mechanical information and determines the criterion for spatially separated motor groups to stay synchronized. Our results generalize the notion of persistence in passive, Brownian filaments to active, cross-linked filaments. 
 \end{abstract}
\maketitle
\end{titlepage}

\begin{figure}
\begin{center}
\includegraphics [width=\columnwidth] {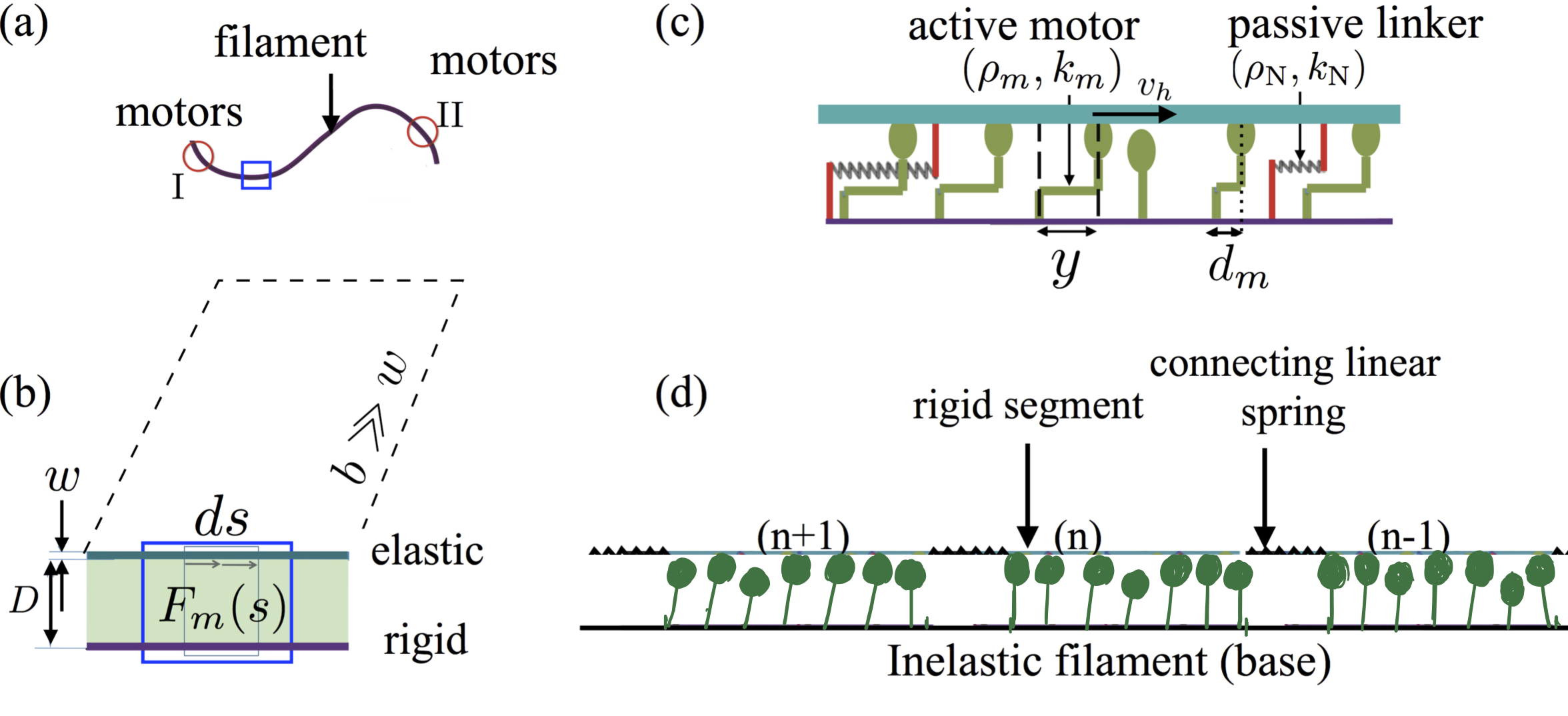}
\caption{\label{fig:1}(a) An active,  motor-filament composite with spatially separated motor aggregates,
(I) and (II).  The softer the intervening filament segment connecting (I) and (II), the higher the tendency of the motor groups to lose coherence. (b) A close-up of the blue box illustrates the geometry we study. The continuous, ordered composite filament is comprised of a thin sheet of width $w$ and lateral width $b \gg w$ that is a distance $D$ above a rigid substrate. This distance is spanned by active motors generating an active force density $F_{m}$. (c) Blow-up of the blue box in (b) illuminates the discrete elements comprising the continuous filament. Shown are the passive linkers 
(N, red) and active motors (subscript $m$, green) spanning the distance $D$. Motors attach with a pre-extension $d_{m}$ and then move along the upper filament with speed $v_{h}$ relative to their base. Motor extension is $y$ and the local filament displacement is $U$. (d) The computational version of the composite has 80 ($1\leq n \leq 80$) rigid segments connected by linear springs of stiffness $K$. Motors (green) are active linearly springs (spring  constant $k_m$) which attach with probability $p_{\mathrm{on}}$ and detach with load dependent probability $p_{\mathrm{off}}$. The free end (last segment) is subjected to a periodic displacement while the first segment is clamped.} 
\label{fig:1}
\end{center}
\end{figure}

\section{Introduction}

Semi-flexible and flexible filaments interacting with active molecular motors arise naturally as constituents of the cell and its organelles \cite{Howard, Witman, Machin, Brokaw_1, Mukundan} as well as in reconstituted in-vitro motor assays \cite{Dogic-2011, Dogic-2012}.  An example is the structurally ordered eukaryotic flagella and cilia. These conserved structures with ubiquitous functions stemming from already being present in the last eukaryote common ancestor \cite{Chlamy},  provide a great opportunity to study the integration of mechanics and control at the cell level. Cilia are comprised of almost inextensible microtubule filaments,  active ATPase dynein motor proteins and passive elastic nexins which in concert deform (bend) and oscillate with well defined wavelengths and frequencies.  
The dynamics of this and of similar active and activated synthetic systems, are controlled by localized forces due to active motors mediated by passive filament elasticity, geometric constraints and noise. 

To understand the onset and persistence of such coordinated deformation, a variety of models with varying degrees of complexity have been proposed. Most of these models \cite{Camalet, Howard_2,  Everaers, Heussinger, Grill, Vilfan_1, Hilfinger}, assume the motor activated filaments to be inextensible -- that is, local extensional strains imposed on the filament can propagate an infinite distance along its contour.  Any propensity to bend or elongate however removes this divergence.
Indeed, in a passive context \cite{Heussinger}, it has been demonstrated that when filament bundles are bent, the shearing forces between them are mediated by extensibility, leading to a characteristic scale over which mechanical information is transmitted.  A recent study also indicates significant influence of finite extensibility of actin filaments in inter-filament friction~\cite{ward2015}.
In active contexts such as for eukaryotic flagella, models ignoring extensibility predict that the wavelengths characterizing the beating filament scale with the flagella length \cite{Camalet}. Experimental evidence however indicates that actual wavelengths  are self-limiting \cite{Witman, Howard_2}  even as the flagella themselves range from tens of microns to nearly a centimeter long.  This strongly suggests a finite elasticity dependent length scale for propagation of mechanical information along the flagella. 

Figure \ref{fig:1}(a) depicts an active composite filament of fixed width comprised of two filaments cross-linked by active motor groups. Consider now the interaction between two spatially separated, distinct, motor patches (I, II). For inextensible filaments, mechanical activity by group (I) leads to local shear and slide; these mechanical signals are transmitted over arbitrarily long distances and can thus be detected by the motor group at (II) however large the separation. For extensible filaments - even when the extensibility is weak - the interplay between sliding deformation and filament elongation modes along the contour eventually results in degradation of mechanical signals transmitted along the filament. This decay of mechanical information limits the range over which motor coordination may occur. In short, 
softness leads to de-coherence and stiffness ensures coherence.  Additionally, kinetics of motor activity set by ATP hydrolysis rates can couple back to effective  extensibility by introducing shear stiffening thus enhancing or disrupting coherence between the  separated motor groups.

Here, we address two questions. First, what sets the length scale over which elastic deformations persist in such systems? Second,  how do elasticity and motor kinetics control the spatial coordination and coherence of collective dynamics? Recognizing that motor groups interacting with an assumed infinitely-stiff filament will immediately coordinate their dynamics,  we hypothesize that elasticity - be it extensional elasticity, active elasticity or shear softening - can mediate interactions between spatially separated regions of the filament. Building on this allows us to identify both passive and active ingredients that control transmission and degradation of mechanical information.   We then analyze a minimal, mean-field  theory in the limit of weak noise and focusing on extensional systems to extract the length scale(s) over which actively generated strains persist.  Finally, we corroborate our theory using discrete simulations that extend our results to moderately stiff and/or noisy systems. 

\section{Analytical model and equations}

We begin by defining the geometry of the model composite active filament illustrated in Fig.\ref{fig:1}(b). A weakly extensible filament of length $\ell$, thickness $w \ll \ell$, Young's modulus $E$ and lateral extent $b$ is held a fixed distance $D$ ($w \leq D \ll \ell $)  above a rigid filament, also of thickness $w$. Connecting the two filaments and maintaining $D$ constant are passive, linearly elastic permanent cross-linkers spaced uniformly with areal density $\rho_{\mathrm{N}}$ and spring stiffness $k_{\mathrm{N}}$. The passive elasticity of the composite is a combination of (i) weak extensibility of the filament as characterized by its stretching/extensional modulus $K$, and (ii) the shear modulus per unit width, $G$ due to the permanent cross-linkers. Consistent with the previously studied cross-linked railway track model~\cite{Everaers}, we define $K=Ew$ and $G=\rho_{\mathrm{N}}k_{\mathrm{N}}$.  

Active forces are generated by the binding of uniformly spaced, unidirectional molecular motors having areal density $\rho_m$ to the upper filament. The base of the motors (tails) are grafted to the lower filament while the heads attach periodically to the upper filament thus exerting a force. Each motor is characterized by an internal variable -- its extension -- that serves as a indicator of how much it is stretched. Motors attach in a pre-strained state with an (initial) extension $d_m$. Once attached, in order to relax the pre-stress, the head slides along the filament, resulting in a changing motor extension $y \neq d_{m}$ and thus exerts an active force. Following previous work \cite{Grill}, we treat the active motors as linear springs with spring constant $k_{m}$, and
assume that motors are characterized by a linear force-velocity relationship with 
stall force $F_{s}$ and a zero-load speed $v_{o}$. This then yields for the speed of the motor head  (SI-$\S$IA), 
\begin{equation}
 \partial_t y =  \partial_t U + v_{o}\left(1-{k_{m}y /F_{s}}\right)
\end{equation}
where $U(s,t)$ is the displacement of the material point on the filament where the motor is attached,  $s$ is the arc-length variable parametrizing location of material points along the filament and $t$ the time. 
For inter-link and inter-motor spacings much smaller than the filament length and  when $b \gg {\mathrm{max}}(w,D)$ a continuum, one-dimensional description of the passively cross-linked, motor-filament aggregate is appropriate
with the activity now treated as arising from an internally distributed force density per motor, $F_{m}$ (Fig. 1b). This density is related to the average fraction of attached motors, $N$ via
\begin{equation}
F_m = k_{m}N\langle y \rangle.
\label{fm}
\end{equation}
Since most motor-filament interactions occur in the over-damped limit with the total sum of forces on the filament vanishing, stress balance yields
\begin{equation}
\partial_s(K \partial_s U) - G U + \rho_{m}F_{m} = 0.
\label{eq:balance}
\end{equation}
In the absence of activity $F_m=0$;  (\ref{eq:balance}) then predicts that local perturbations in $U$ decay exponentially with a length scale $\ell_E \equiv (K/G)^{1\over 2} =  ({Ew / \rho_{\mathrm{N}}k_{\mathrm{N}}})^{1 \over 2}$. 

To complete equations (1)-(3), we need to determine how the the fraction of attached motors,  $N$, evolves.  In the mean-field limit at high motor densities (large $\rho_{m}$) fluctuations of $N$ about the mean are negligible with the attachment and detachment motor fluxes simply related to the mean attachment  $\on$ and detachment $\off$ rates.
Let $\delta_{m}$ be the characteristic motor extension at which attached motors detach; the potential energy that is lost when a motor detaches is then $\mathcal {E} = k_m \delta_m^2/k_BT$, with $T$ being the temperature. Ignoring motor diffusion, we ensemble average the 
microscopic balance equations (SI-$\S$1B \& 1C)
to obtain evolution equations for $N$ and for the scaled extension, $Y \equiv \langle y \rangle/\delta_{m}$,
 \begin{eqnarray}
\partial_t N &=& {\omega}^{o}_{\mathrm{on}} (1-N)- {\omega}_{\mathrm{off}}N \\
\partial_t Y  &=&   \frac{\partial_t {U}}{\delta_{m}} + {\mathcal{A}}_{1} {{\omega}^{o}_{\mathrm{off}}} \left({ {\mathcal{A}}_{2} }-Y\right) + {{\omega}^{o}_{\mathrm{on}}} 
{{{\mathcal{A}}_{3} - Y} \over {N/(1-N)}}.
\end{eqnarray}
The first term on the right hand side of the equation (5) is the passive convection of the motor head due to filament extension, and the second and  third terms reflect the active interaction between the motor head and the filament. 
Without loss of generality, we treat the mean attachment rate $\omega^{o}_{\mathrm{on}}$ as constant, while the mean detachment rate  $\omega_{\mathrm{off}} =\omega_{\mathrm{off}}^{o} {\mathcal F}({\mathcal{E}},\langle y \rangle/\delta_{m})$ is allowed to vary with motor extension $Y$ and the energy scale $\mathcal {E}$ through the function $\mathcal{F}$. 
Here, we choose ${\mathcal{F}}({\mathcal{E}},\langle y \rangle/\delta_{m}) 
= {\mathrm{cosh}}({\mathcal{E}}\langle y \rangle/\delta_{m})$. Choosing a different functional form does not introduce qualitative differences. Equations  (2)-(5) then involve three dimensionless parameters $\mathcal{A}_1 \equiv v_{o}k_{m}/(\omega^{o}_{\mathrm{off}}F_{s}) $, ${\mathcal{A}}_{2} \equiv F_{s} /(k_{m} \delta_{m})$ and $\mathcal{A}_{3} \equiv {d_{m} / \delta_{m}}$ 
apart from the purely kinetics based motor duty ratio that is related to $\Psi \equiv \off^o/\on$. To ascertain the role of pre-strain, we note that when $\mathcal{A}_1 \ll 1$  the motor extension is dominated by the pre-strain $d_m$; conversely, 
when $\mathcal{A}_3 \ll 1$, pre-strain has negligible influence. 

Our model complements and differs from previous attempts in a few important ways. First, consistent with experiments suggesting that bond failure is more naturally dependent on the extension and only weakly on the rate of extension, we have chosen ${{\omega}^{o}_{\mathrm{off}}}$ to depend on motor extension \cite{Vilfan_1, Grill} and not the rate of extension \cite{Hilfinger}. Second, non-linear coupling between passive and active deformations in (3)-(5) distinguishes our model from previous studies of motor mediated bending of filaments \cite{Hilfinger}. Finally, our model filament is weakly extensible, bolstered by recent experimental evidence \cite{Heussinger,Howard_2,ward2015}.

\section{Persistence lengths in weakly elastic systems}

As a first step towards understanding persistence of strains in long elastic composite filaments and the evolution of global strain fields, 
we analyze conditions for the emergence of localized strains in a stationary filament. Referring to the blue box in Fig 1(b),  we focus on a small
localized patch of motors  interacting with and animating a fragment of length $\ell_{s} \ll (K/G)^{1 \over 2}$. This fragment, which strains negligibly and is therefore rigid to leading order,  then tries to move relative to its neighbors on either side.
The ensuing dynamics can be mapped to that of an animated rigid segment working against an effective spring with passive and active components. We lump these contributions using an effective spring constant $K_{s}$ and write for the displacement of the segment
\begin{equation}
- K_{s} U  + \rho_{m} k_{m} \delta_{m} N Y = 0.
\label{eq:short}
\end{equation}
For the physically relevant condition $K_{s} >0$, global filament translation is prevented, and the non-linear equations (4)-(\ref{eq:short}) admit two stable solutions -- a stable, stationary constant displacement of the rigid filament  or stable time-periodic, oscillations. 
The static state to found to be linearly unstable  (SI-$\S$ IIA \& IIB) to oscillatory states via supercritical Hopf-Poincare bifurcations similar to that predicted in models for flagellar \cite{Hilfinger} and  spindle oscillations \cite{Grill}. 
We have thus demonstrated  that  {\em localized steady or oscillatory extensional strains may spontaneously emerge} in a small fragment of a larger active composite. For a filament with infinite elastic modulus $K \rightarrow \infty$, the effect of localized strains is felt everywhere along the filament; this enables spatially separated motor groups to act coherently.  

\begin{figure}
\begin{center}
\includegraphics [width=\columnwidth] {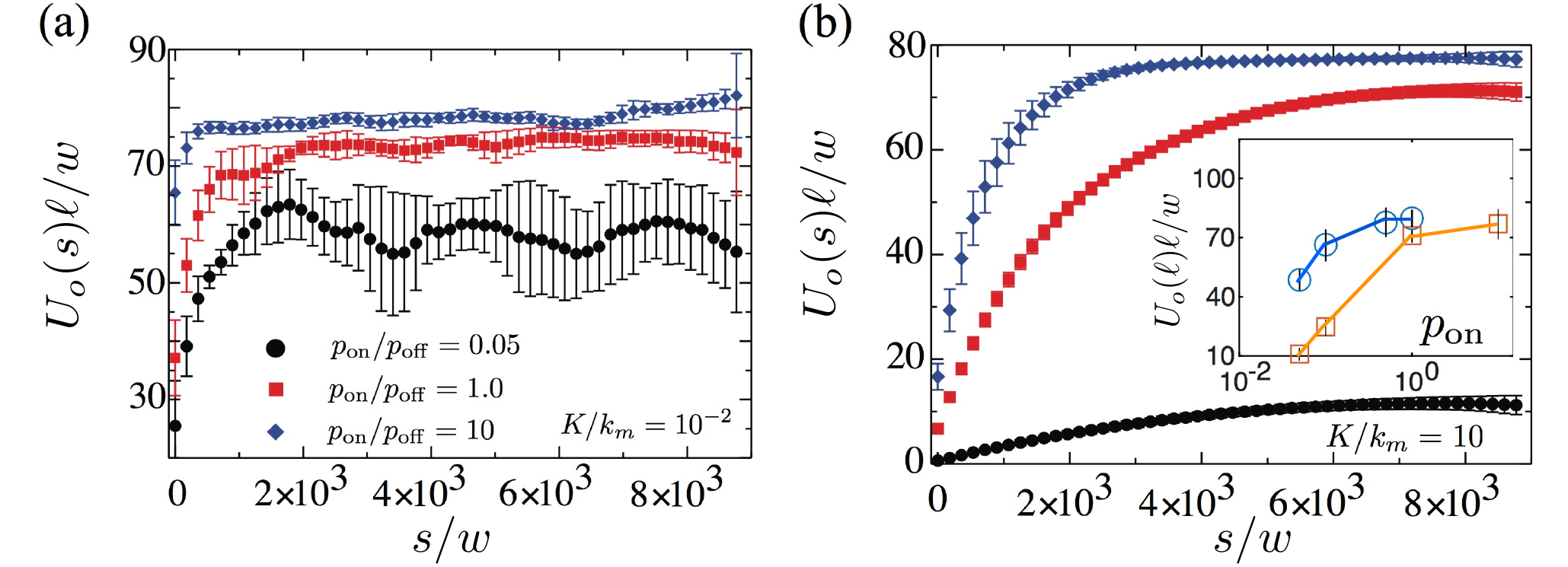} 
\caption{Simulation results: Scaled mean steady extension as a function of position from the clamped end. Vertical bars correspond to the RMS deviation.  (a) Extensions are uncorrelated for soft assemblies ($K/k_{m} = 0.01$) with domains fluctuating independently.  (b) For stiff filaments ($K/k_{m} = 10$),  extensions are highly correlated but without a decay length. (inset) The extension at the free end as a function of attachment probability $p_{\mathrm{on}}$ for soft ($K/k_{m} = 0.01$, circles, online-blue) and  stiff ($K/k_{m} = 10$, squares, online-red) filaments.}
\label{fig2}
\end{center}
\end{figure}

Starting from this limit of perfect coherence, we now analyze how weak elasticity modulates the decay of strains, introduces a persistence length and disrupts coherent behavior.
We begin by identifying a suitable base state that sets the strain field over the length of the composite filament. 
Stationary, steady filament displacement, $U_{0}$,  attached motor fraction, $N_{0}$, and motor strain, $Y_0$ fields are obtained from (\ref{eq:balance})-(5) 
\begin{eqnarray}
 \label{eq:steady1}
 K \partial_{ss} U_{0} - G U_{0} + G_{\mathrm{act}} \delta_{m} N_{0} Y_{0} = 0, \\
N_{0}= (1+ \Psi {\mathcal F}_{0})^{-1}, \\
Y_{0} = ({{\mathcal{A}}_{1} {\mathcal{A}}_{2}} + A_{3} {\mathcal{F}}_{0}) ({ {\mathcal{A}}_{1} + {\mathcal{F}}_{0}})^{-1},
\label{eq:steady}
\end{eqnarray}
where $G_{\mathrm{act}}=\rho_mk_m$
is the active analogue of the passive shear modulus, $G$
and ${\mathcal{F}}_{0} = {\mathcal{F}}(Y_{0})$.  We choose boundary conditions consistent with $s=0$ held fixed ($U_{0}(0)=0$) and $s=\ell$ left free ($\partial_s U_{0}(\ell)=0$), and solve 
(\ref{eq:steady1}) to obtain
\begin{equation}
U_{0}  =  {{G_{\mathrm{act}}} \delta_{m} N_{0} Y_{0} \over {G}}  \left( 1 - \alpha \: e^{{s / \ell_{E}}}
-   \left(1 - \alpha \right) \: e^{-{s / \ell_{E}}}\right)
\label{eq:u0}
\end{equation}
where $\alpha \equiv  e^{-2{\ell / \ell_{E}}}/(1 + e^{-2{\ell / \ell_{E}}})$ and
the decay length $\ell_E = \sqrt{K/G}$.
While surprising at first sight, the expression for the decay length is rationalized by recognizing that attached motors sense only strain rates and not the actual strain;  and in steady extension these strains rates are zero. 
The decay  length is modified substantially when the motors are in rigor. Setting the attachment and detachment rates to zero as is appropriate, we find that the decay length is now given by $ \sqrt{{K / ({G_{\mathrm{act}}N_{0} + G}})}$ (SI-$\S$IIIA, \cite{Inhomosoln}) and thus strongly influenced by the fraction of attached motors at rigor.

We next analyze the decay of oscillations strains in a filament.  To do so, we modify the boundary condition at $s=\ell$ to enable imposed oscillations there, while still respecting the constraints that lead to (\ref{eq:u0}). This is achieved by subjecting the free end to a small amplitude oscillatory displacement with frequency $\omega$ and amplitude $\epsilon U_{I} \ll U_{0}(\ell)$ with $\epsilon \ll 1$.  
The boundary conditions then take the form $U(0,t) = 0$ and $U(\ell,t)=U_{0}(\ell) + \epsilon U_{\mathrm{I}}{\mathrm{Real}}[e^{i\omega t}]$.
To study the frequency-locked (adiabatically slaved) response of the filament to this localized imposed oscillation, we 
write $(U, N,Y) = (U_{0}, N_{0},Y_{0})  + e^{i \omega t} \epsilon (\hat{U}, \hat{N},\hat{Y})$ and substitute this form in equations (2)-(5). Retaining terms to $O(\epsilon)$, we find
\begin{eqnarray}
K\:\partial_{ss} \hat{U} -  G \hat{U} - G_{\mathrm{act}} \chi \hat{U} &=&0\\
\hat{N} i \omega  + \hat{N} \omega_{\mathrm{on}}^{o} (1 + \Psi {{\mathcal F}_{o}})  + \omega_{\mathrm{on}}^{o}  \Psi N_{o}
{{\mathcal F}'_{o}} \hat{Y} &=& 0\\
{{ i \hat{Y}\omega} \over \omega_{\mathrm{on}}^{o}} +   \hat{Y} \Psi {\mathcal{A}}_{1}
{{\mathcal F}_{o}} 
  + ({\mathcal{A}}_{3} - Y_{0}) {\hat{N} \over N_{0}^{2}} &=& {{ i \omega \hat{U}}  \over {\delta_{m}\omega_{\mathrm{on}}^{o} }}.
\label{eq:oscillatory}
\end{eqnarray}
where the compliance function $\chi = - \delta_{m} (N_{0} \hat{Y} + Y_{0}\hat{N}) / \hat{U}$ determines  the linear  viscoelasticity of the composite filament and depends on motor characteristics such as the stall force $F_{s}$,  and free velocity, $v_{0}$  and the duty ratio $\Psi$.  Examining the function, we deduce that  the frequency-locked response is possible only when Real[$\chi$] is negative. 
The criterion for non-trivial solutions to exist (see SI-$\S$ III) 
%$(G +  G_{\mathrm{act}} \mathrm{Real}[\chi] - K/ \lambda^{2})^2 +G^{2}_{\mathrm{act}} \mathrm{Im}^{2}[\chi]=0$.
then provides the expression for the {\em effective} persistence length scale:
%$\lambda_{\mathrm{E}} \equiv {\mathrm{Real}}[\lambda] $:
\begin{equation}
\lambda_{\mathrm{E}} = \sqrt{2}\: \ell_{E} \: \left(\sqrt{1 + \beta^{2} |\chi|^{2} + 2 \beta {\mathrm{Real}[\chi]}} + 1 + \beta \:{\mathrm{Real}[\chi]}\right)^{-{1 \over 2}}
\label{eq:persistence}
\end{equation}
with
$\beta \equiv G_{\mathrm{act}}/G$, 
controlling the relative importance of active to passive shear stiffening. 
In the absence of activity, the decay length reduces to its value for a passive filament - i.e, $\lambda_{\mathrm{E}}  = \ell_{E}$. Expanding (14) for weak activity ($\beta \ll 1$) and strong activity ($\beta \gg 1)$ (\cite{Asymptotics_1}, \cite{Asymptotics_2}, SI-$\S$III D \& E) shows that {\em activity can either enhance or reduce persistence length due to transient shear stiffening provided by the motors} -- an effect that depends strongly on the frequency $\omega$ and the ratio $G_{\mathrm{act}}/K$. 

\begin{figure}
\begin{center}
\includegraphics [width=\columnwidth] {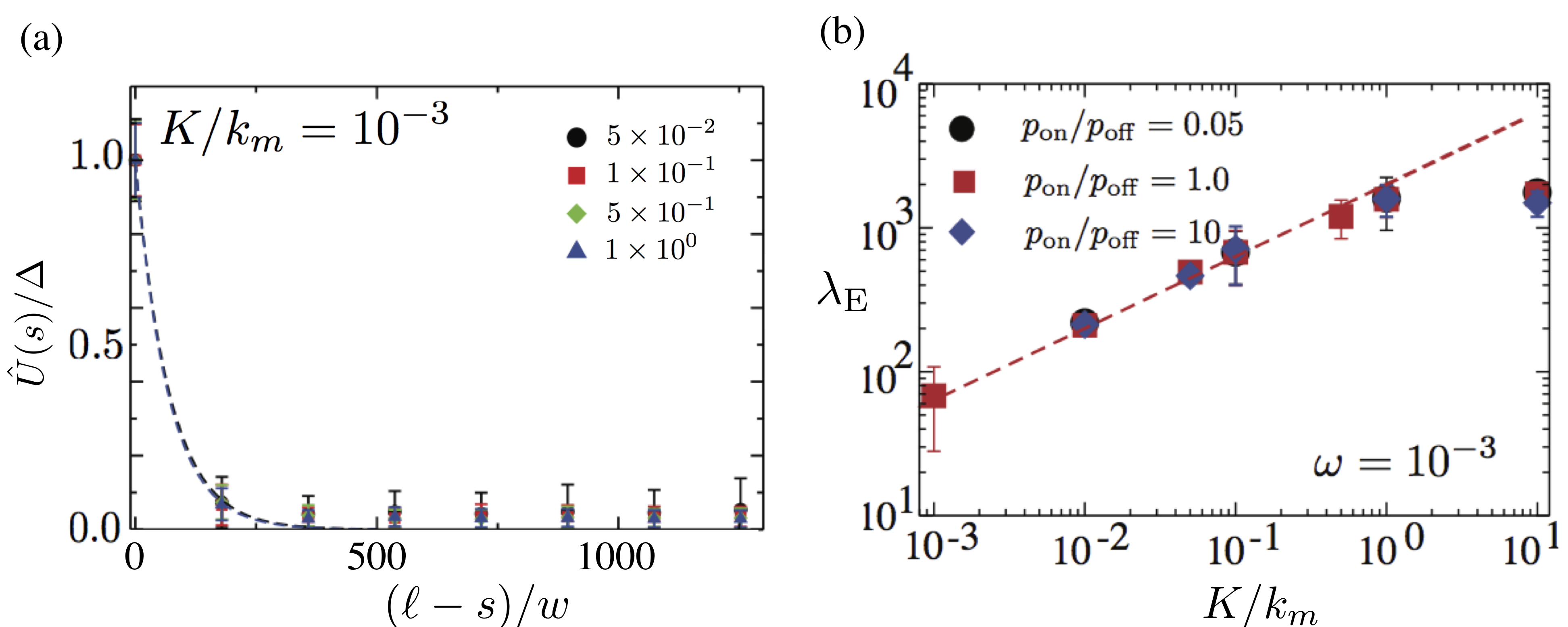}
\caption{(a) Averaged amplitude of the oscillations in the strain field along the filament a function in the limit of extreme softness, $K/k_{m}=10^{-3}$. An exponential decay length is indicated with variances shown as vertical bars. The area near the clamped end is not shown for clarity. (b) Decay length of the amplitude in the propagated oscillatory strain as a function of $K/k_m$ for various $p_{\mathrm{on}}/p_{\mathrm{off}}$. System size effects result in the saturation of  the decay length for $K/k_{m} \gg 1$.
\label{fig3}
}
\end{center}
\end{figure}

\section{Persistence lengths in noisy systems}

The mean field prediction (14) ignores thermal fluctuations in the active force density $F_{m}$,  motor noise due to discreteness of binding and unbinding events, fluctuations in the number of attached motors and synergistic interactions between attachment probabilities  and filament elasticity. To highlight the effects of  motor noise, activity and moderate softness on the persistence length, we study a discrete model filament-motor composite (see Fig.1(d)), using Brownian-MPC simulations (\cite{Simulation_1, Simulation_2}, SI-$\S$IVA) with $G=0$ with $K>0$ and $G_{\mathrm{act}} > 0$. Results are summarized in Figures 2 and 3.

We find that 
the decay of steady, motor-mediated, localized extensions  -- Fig. \ref{fig2}(a) and Fig. \ref{fig2}(b) --  is controlled by $K/k_m$ and the ratio of detachment to attachment probabilities, $p_{\textrm{off}}/p_{\textrm{on}}$,  \cite{Simulation_2} for both stiff ($K/k_{m} = 10$) and soft ($K/k_{m} = 10^{-2}$) filaments. The mean field prediction (\ref{eq:u0}) for $G=0$,  
$U_{0}(s) \propto \rho_{m}\left({k_{m}/ K}\right)
 s (2\ell -s)$ is qualitatively consistent with the simulation results for $p_{\textrm{off}}/p_{\textrm{on}} \sim 1$ with deviations seen for large contrasts between the probabilities. Increasing the attachment probability results in a larger fraction of attached motors with sharper gradients near $s=0$ and flatter profiles near $\ell$ consistent with analytical predictions (\cite{Inhomosoln}).   The extension at $s=\ell$ increases with increasing $p_{\textrm{on}}/p_{\textrm{off}}$ consistent with higher levels of motor attachment \cite{dutyratio} yielding longer extensions. Motor noise results in large fluctuations in the mean extension, noticeable especially for the soft filaments and also results in uncorrelated spatial domains in extension \cite{Fluctuations}.  The decay of localized oscillatory strains was also analyzed by introducing an localized oscillation $U(\ell,t)=\Delta \sin(\omega t)$ about the pre-strained state $U_0(s)$ with $\omega$ chosen to be smaller than the turnover frequency of motors so as to maintain the frequency-locked response. The extensional field $\hat{U}(s)$ decays exponentially as seen in fig. \ref{fig3}(a).
Plotting the values of $\lambda_E$ for various values of $K/k_{m}$ in Fig.\ref{fig3}(b), we find that over the range of $p_{\textrm{off}}/p_{\textrm{on}}$ and $\omega$ we studied, $\lambda_E$ has a rather weak dependence on motor activity. 
Surprisingly, the decay length  follows the prediction of (14) 
$\lambda_E \sim \sqrt{2 K/k_m}$ very well over a wide range of filament softness.
We note that very soft filaments,  motor noise causes fluctuations in both amplitude and phase with sections of the filament demonstrating significantly de-correlated response.

To summarize, with motor properties held fixed, the decay length sets a finite range of correlated activity that is relevant for naturally ordered active matter such as eukaryotic flagella~\cite{Woolley}. Our results for the persistence length can be tested experimentally by examining  spatially separated motor bundles interacting with the same filament  \cite{Dogic-2011,Dogic-2012}.  In ordered systems such as flagella and muscle, experiments probing the maximum flagellar wavelengths and maximum muscle fiber lengths will serve to test our theory. A simple estimate approximating the flagellum as an actively driven filament~\cite{Flagella, Howard_1, Howard_2, Lindemann, Camalet} yields $\ell_{E} \in [80-200]$ $\mu$m - a range within physically observed lengths. 
Our results are also relevant to understanding the role of extensibility in soft  filaments subject to follower forces \cite{Chellakot_1} as well as to actively deformed two and three dimensional motor-filament networks \cite{FN1, FN2} where motor properties may vary in response to the network elasticity.


\begin{thebibliography} {32}
\bibitem{Inhomosoln}
For a spatially varying total (active) motor density, the equation for the extension is
$
\ell^{2}_{\mathrm{E}}d_{ss} U - U +  C^{*}\phi(s)= 0 $ where $C^{*} \sim k_{m} \rho_{m}/G$ depends linearly on
$\rho_{m}$, and $\phi(s)$  encodes the spatial variation. The solution that satisfies the clamped-free boundary conditions is
$U_{o}(s) = 2A\left[{\mathrm{sinh}}(s/\ell_{\mathrm{E}})\right] + \ell_{\mathrm{E}} C^{*}\int_{0}^{s} \phi(s^{*})\:{\mathrm{sinh}}({{s^*-s}\over \ell_{\mathrm{E}}})\: ds^{*}$. Here, $A = {1 \over 2}  {\ell_{\mathrm{E}}} C^{*} \int_{0}^{\ell} \phi(s^{*})  \left[{{\mathrm{cosh}}({{s^{*}-\ell} \over \ell_{\mathrm{E}}})/{\mathrm{cosh}}({\ell \over \ell_{\mathrm{E}}})}\right] \:ds^{*}$.
\bibitem{Asymptotics_1}
At fixed frequency, $\omega$ and for both strong and weak activity in the presence of  
permanent linkers, the persistence length is controlled by $G_{\mathrm{act}}/K$. For weak activity, $\beta \ll 1$, 
$\lambda_{\mathrm{E}}\: \approx \ell_{E} \left(1 - {1\over 2} \beta {\mathcal{R}} - {1 \over 8}\beta^{2}({\mathcal{I}}^{2} - 3 {\mathcal{R}}^{2}) \right) + O(\beta^{3})$. For ${\mathcal{R}} > 0$, the $O(\beta)$ term decreases $\lambda_E$ while the $O(\beta^{2})$ term can either enhance or reduce the persistence length.  For strong activity, $\beta  \gg 1$,  $\lambda_{\mathrm{E}} \:\approx\sqrt{2(K/ G_{\mathrm{act}})} ({\mathcal{R}} + |\chi |)^{-{1 \over 2}}$ with actively driven shear stiffening dominating.
\bibitem{Asymptotics_2}
At low frequency $\omega$, motors undergo multiple mechanochemical cycles before they feel the imposed oscillation.
At high frequency, attached motors encounter several cycles of imposed oscillations before detaching.   
In the low frequency limit when  ${\mathcal{A}}_{1} = 0$ 
$
{{\lambda_{\mathrm{E}}} \over {\ell_{E}}}  \approx 1 - {1 \over 8} ({\omega^{2} / \omega_{\mathrm{off}}^{2}}) 
\{ {\beta^{2} } (1-\gamma + \gamma \tau)^{2} + 4 {\beta} (1-\gamma + \gamma \tau^{2}) \}
$
so that to leading order, activity decreases the persistence length.
For high  $\omega$, $  
{{\lambda_{\mathrm{E}}} \over ({\ell_{E} / \sqrt{1+\beta}})}  \approx 1+{1 \over 8} ({\omega_{\mathrm{off}} /\omega})^{2}\{4 \beta(1-\gamma + \gamma/\tau^{2}) 
-{\beta^{2} \over 1+\beta} (1-\gamma + \gamma/\tau)^{2}$.
When $\omega \rightarrow \infty$, $\lambda_{\mathrm{E}} \rightarrow \ell_{E}/\sqrt{1+ \beta} = \sqrt{K/({G+ G_{\mathrm{act}}})}$.
\bibitem{Simulation_1}
The  computational discrete filament comprises of 50 equal rigid segments 
(width $w$, length $\ell=80 w$),  connected to each other by linear springs of (stiffness  $K$ with zero rest length). The first segment is connected to a rigid wall (at $s=0$) while the other end of the composite ($s=\ell$) is kept free. The active force on the filament is imposed by 80 motors per segment, modeled as linear springs with stiffness $k_m$, with one end (tail) permanently grafted to a rigid base and the other end (head) attaches to the filament with a  $p_{\textrm{on}}$. The motor attachment occurs at zero pre-strain i.e., $d_m=0$. Once attached, the motor head slides on the segment which leads to a motor extension ($y$) and a corresponding motor force.
\bibitem{Simulation_2}
The attachment probability $p_{\mathrm{on}}$  is constant. The detachment probability $p_{\mathrm{off}}$ is load dependent.
To emphasize the role of motor noise, we do not incorporate permanent passive cross links.
\bibitem{dutyratio}
If  $t_{d}$  is the average time that a motor stays detached, then $p_{\mathrm{on}} \sim t_{d}^{-1} \sim  \omega_{\mathrm{on}}$.
\bibitem{Fluctuations}
Variations in the mean extension (shown as vertical bars) are more correlated for stiffer filaments than for softer filaments. The fluctuations in extension decrease with increase in $p_{\textrm{on}}$ due to larger fraction of attached motors.
\bibitem{Flagella}
Consider a model flagellum as a composite filament driven by uniformly spaced motors with $w$, $D$ and $E$ being the radius, inter-filament spacing and microtubule Young's modulus. Let  $k_{\mathrm{N}}$ and $\rho_{\mathrm{N}}$ be the nexin stiffness (per length) and nexin density. From literature,  $w \sim 20$ nm, $D \sim 40$ nm \cite{Howard_1}, $E \sim 1.2$ GPa \cite{Howard_1,Howard_2},  $k_{\mathrm{N}} \sim 16-100$ pN $\mu$m$^{-1}$, $k_{m} = 10^{-3}$ N/m \cite{Lindemann}, $\rho_{m} \sim 10^{8}$ m$^{-1}$ and $w \rho_{\mathrm{N}} \sim 10^{5}-10^7$ m$^{-1}$ \cite{Lindemann, Camalet}. We find the frequency dependent $\ell_{E} \in [10-500]$ $\mu$m ands thus in the range of physically relevant length scales. The lower limit is with all dyneins and nexins attached;  the upper bound is no dyneins attached.  With bending, the persistence length to initiate oscillations   $ \in 80-200$ $\mu$m.
\end{thebibliography}
\end{document}